\documentclass[nologo,url,11pt,a4paper]{ETHpaper}
\usepackage{graphicx}                  
\usepackage{amsmath, amssymb, amsfonts}
\usepackage{dcolumn}
\usepackage[usenames,dvipsnames]{xcolor}

\begin{document}
\title{The Dynamics of Emotions in Online Interaction}

\titlealternative{The Dynamics of Emotions in Online Interaction. Submitted on May 12.}
\author{David Garcia, Pavlin Mavrodiev, Frank  Schweitzer}

\author{David Garcia$^{1*}$, Arvid Kappas$^{2}$, Dennis K\"uster$^{2}$, Frank Schweitzer$^{1}$}

\address{$^{1}$Chair of Systems Design, ETH  Zurich, Weinbergstrasse 56/58, 8092 Zurich,  Switzerland \\
$^{2}$ Jacobs University Bremen,  Campus Ring 1, 28759 Bremen, Germany\\
$^{*}$ dgarcia@ethz.ch }

\date{May 12, 2016}

\www{\url{http://www.sg.ethz.ch}}

\maketitle
\centerline{May 12, 2016}

\begin{abstract}
We study the changes in emotional states induced by reading and participating in online discussions, empirically testing a computational model of online emotional interaction. 
Using principles of dynamical systems, we quantify changes in valence and arousal through subjective reports, as recorded in three independent studies including 207 participants (110 female). 
In the context of online discussions, the dynamics of valence and arousal are composed of two forces: an internal relaxation towards baseline values independent of the emotional charge of the discussion,  and  a driving force of emotional states that depends on the content of the discussion. 
The dynamics of valence show the existence of positive and negative tendencies, while arousal increases when reading emotional content regardless of its polarity. 
The tendency of participants to take part in the discussion increases with positive arousal.
When participating in an online discussion, the content of participants' expression depends on their valence,
and their arousal  significantly decreases afterwards as a regulation mechanism. 
We illustrate how these results allow the design of agent-based models  to reproduce and analyze emotions in online communities.
Our work empirically validates the microdynamics of a model of online collective emotions, bridging online data analysis  with  research in the laboratory.
\end{abstract}

\section{Introduction}

Intergroup emotions theory, based on appraisal theory \cite{Lazarus1991,Smith1985,Scherer2001} and the social identity perspective \cite{Turner1987} suggests that dynamic variability of emotions over time can be found not only at the individual level but also at the level of group and collective emotions \cite{Smith2015,Scheve2013}. More specifically, emotional experience has been found to elicit the social sharing of emotion in the construction of emotional climates \cite{Rime2009,Rime2007}, via processes that are likely to involve group identification \cite{Smith2007}, emotion contagion \cite{Weisbuch2008}, conformity with group norms \cite{Turner1987}, and social or  group-based regulation of emotion \cite{Kappas2013,Smith2015}, to name a few. It is likely that such processes are not limited to face-to-face encounters, in particular since many of the sources 
of over-time variability of group-based emotions  \cite{Smith2015} and collective emotions  \cite{Scheve2013} may be at least partially of an informational nature.
Participatory online communities, such as social networking sites or discussion fora, provide a thriving medium for the emergence of such collective emotional phenomena.

Collective emotions are defined as states of an community in which a large number of individuals share one or more emotional states \cite{Scheve2013,Kappas2013}. While the differences between face-to-face and online interaction are evident, collective emotions can also emerge in online communities \cite{Schweitzer2010,Scheve2014}, a topic that is receiving rising attention in the literature (see chapters 25 to 28 of \cite{Scheve2014}).
In the field, collective emotions can be found frequently in online communities, including spontaneous fights in political fora \cite{Chmiel2011}, the spreading of Internet memes \cite{Guadagno2013} and political movements \cite{Alvarez2015}, or user-generated content on YouTube that may, under certain circumstances, become viral through the excitement of thousands of users \cite{Shifman2012,Garcia2012}.
For example, in the 2008 Presidential campaign, supporters of the Obama campaign successfully initiated the spreading of the well-known ''Yes We Can'' video that had attracted over 20 million views by the time of the democratic nomination \cite{Wallsten2010}. 
In such cases, there is an intriguing interplay of collective emotions online and emotional behavior offline.

While collective emotions also appear in offline situations, the case of online interaction has recently attracted a lot of attention for three reasons: i) Internet access has spread rapidly over the past decade \cite{Duggan2013}, ii) its usage has noticeably diversified \cite{Deursen2014}, and iii) the traces left on publicly accessible posts and messages allow quantitative analyses of unprecedented size and resolution.
Examples of the latter are recent insights on collective emotions that have been derived from forum discussions \cite{Chmiel2011}, real-time chatroom conversations \cite{Garas2012}, and Twitter messages \cite{Thelwall2011}.
The role of collective emotions in communication is, obviously, not entirely new to the Internet. 
In fact, communication sciences have been studying mass-mediated one-to-many communication for decades, particularly in the context of print media, radio, and television \cite{Roberts1981}. 
Collective emotions in offline scenarios are related to the  concept of emotional contagion \cite{Hatfield1994}, 
which is likely related to several communicative processes, such as dyadic mimicry
\cite{Hess2013}. 
Online interaction is still mainly based on textual communication, which has been shown to create interpersonal emotional interaction in real-time chats \cite{Garas2012}.
Furthermore, controlling social media interaction by its emotional content creates small linguistic cues in written emotional expression \cite{Kramer2014}. 
Online written interactions have the power to elicit emotions in the users of online media \cite{Kappas2011}, offering us the chance to combine observational and experimental studies in a unified view of online collective emotions.

Emotional interaction leads to the emergence of collective emotional states that cannot be simplified to behavior observed when individuals react in isolation to emotionally relevant events \cite{Kappas2013}. 
This \emph{micro-macro} gap calls for models that can explain the emergence of collective emotions from social interaction \cite{Kuester2013}.  
An established approach to understand collective social phenomena is agent-based modeling \cite{Schweitzer2003}.
Agent-based models are mathematical formulations that define the states and interactions of agents, allowing the analysis of collective behavior either through simulations or through methods from statistical physics.
The aim of agent-based modeling is not to design a detailed model that includes all possible aspects of the behavior of the agents, but to propose a minimal set of computation rules and dynamics that lead to the observed collective behavior.
A paradigmatic example is Schelling's model \cite{Schelling1971}: an extremely simplified model of spatial mobility that shows how 
social segregation emerges from weak individual biases \cite{Ball2007}.

Agent-based approaches and computational models are still scarce in empirical psychology \cite{Kuppens2010}. 
However, this tradition may be about to change. 
Agent-based models can provide new insights into social psychology \cite{Smith2007}, unifying different models of social interaction into a comprehensive computational representation of emotions that can encompass
a variety of aspects, including identification processes \cite{Tajfel1978} that can take place during online social interaction \cite{Cheung2011}, changes in self-categorization as a group member \cite{Seger2009}, emotional contagion \cite{Hatfield1994,Weisbuch2008}, as well as affective-discoursive patterns that may help to reconnect discourse studies with novel research on affect and emotion \cite{Wetherell2013}. Towards this aim, the modeling in the present paper will be based on the valence and arousal circumplex of core affect \cite{Russell1980,Russell2009,Yik2011}.  
This view is not specific to emotion, but provides some grounding to the application of our techniques in social psychology. 
Regarding emotional dynamics, some works within mathematical psychology provide support for the usefulness of using agent-based modeling in such contexts \cite{Scherer2009}.
For example, coherence in self-evaluation has been modelled as a cellular automaton \cite{Vallacher2002},
in which memories are represented by cells and the attitude of an individual to certain memory evolves depending on its relationship to other memories, leading to the emergence of self-esteem.
Concepts of dynamical systems can also be applied to model emotions, for example explaining fight or flee reactions as bifurcations \cite{Sander2005} in which the emotional state of an organism can sharply change depending on a control parameter.
Furthermore, the principle of Brownian agents \cite{Schweitzer2003} has been proved useful to analyze the temporal evolution of core affect \cite{Kuppens2010}, modeling changes in emotions as equations with a deterministic and a stochastic component.
In addition, agent-based models of emotions are also used in the context of 
momentary subjective well-being \cite{Rutledge2014} and in cue-reward learning \cite{Watanabe2015}.
While these models explore and validate aspects of individual emotions, agent-based models of emotions under social interaction are still to be empirically analyzed.
Our focus on the social aspects of emotions is aimed at
the design of models with potential to explain and reproduce collective emotions.

Agent-based models can be very useful to predict future user behavior, or to design mechanisms that optimize certain aspects of the community.  
But this cannot be achieved if the microdynamics that drive agent actions are not validated beyond computer simulations or analytic results. 
It is relatively simple to design a model that, based on ad hoc assumptions, reproduces any observed macroscopic behavior. 
As part of an interdisciplinary collaboration to understand the dynamics of collective emotions in online communities \cite{Ahn2011}, the Cyberemotions modeling framework \cite{Schweitzer2010} 
was designed to provide generative mechanisms of online collective emotions, explicitly 
avoiding the pitfall of using ad hoc assumptions and implausible dynamics.
This framework allows the creation of models of user emotions under different kinds of online interaction,  linking collective behavior with individual dynamics in the presence of online interaction mechanisms. 
The dynamics of emotions of agents in the Cyberemotions framework are phrased within the psychological theory of core affect \cite{Russell1980}, which provides a unified representation of the kind of emotions we refer to.
While collective behavior can be analyzed through observational studies of large scale datasets of online interaction, the individual dynamics must be empirically tested in experimental studies.

The Cyberemotions modelling framework has been shown useful to understand the conditions that lead to collective emotions in product reviews \cite{Garcia2011} and chats \cite{Garas2012}.  
Following the concept of agent-based theory building in social psychology \cite{Smith2007}, we phrase the agent dynamics that reproduce online collective emotions \cite{Schweitzer2010} as testable hypotheses. 
In this article, we formulate and test those hypotheses against data collected in experiments on emotion dynamics.
We report the results of three independent studies that allow us to quantify emotional changes while reading and writing posts in online fora. 
We measure the functional dependencies between online content and changes in emotional states, to compose a computational model that can simulate the dynamics of emotions under online interaction.

\section{Emotional agents model}

In the Cyberemotions modeling framework \cite{Schweitzer2010}, the emotional state  of an agent is composed of two variables: valence $v_i(t)$, quantifying the degree of pleasure associated with an emotion, and arousal $a_i(t)$, representing the degree of activity associated with the emotion.  
Agents have an expression variable $s_i(t)$ that quantifies their displayed emotions through online posts, which are aggregated with the expression of other agents in the online field of interaction $h$. 
The first assumption of our model is the existence of internal \emph{eigendynamics}
that make emotions relax to an equilibrium state. 
Secondly, the changes in emotions due to social interaction, or \emph{perception dynamics}, are defined as functions that  take as input the values of valence and arousal and the interaction field $h$.
Thirdly, the expression of an agent through $s_i(t)$ is assumed to be triggered by \emph{expression rules} that determine the value of $s_i(t)$ based on the emotional state of the agent, leading to a \emph{feedback of expression} that changes in the emotional state of the agent after expression.
The field variable $h$ takes positive values when online discussions are positively charged, negative when discussions are negatively charged, and $0$ when no clear emotionality is present in online interaction. 

\begin{figure}[ht] 
\centering
\includegraphics[width=0.75\textwidth]{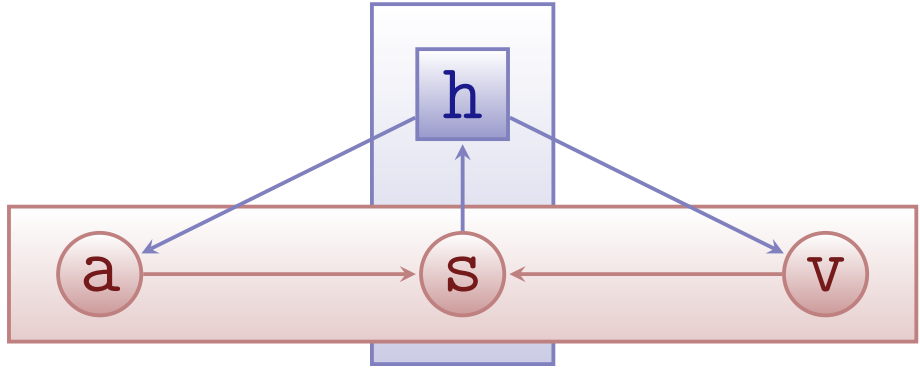}
\caption{Schema of the agent-based model of emotion dynamics \cite{Schweitzer2010,Garcia2011}.
The internal state of an agent is composed of valence $v$ and arousal $a$, and the communication through the online medium is quantified by the field $h$. 
Valence and arousal change according to  a combination of internal \textit{eigendynamics} and \textit{perception dynamics}, the latter  depending on the field $h$. 
Expression $s$ is triggered by \textit{production rules} depending on the emotional state of the agent.
Expression changes the field and leads to a \textit{feedback of expression} that regulates the emotions of the agent. \label{fig:dynSchema} }  \end{figure}

Figure \ref{fig:dynSchema} depicts the main variables of the framework and their influences.

We model the dynamics of emotions as a linear combination with a stochastic component, following the principle of Brownian Agents \cite{Schweitzer2003}:
\begin{eqnarray}
  \frac{\delta v_i(t)}{\delta t} &=& - \gamma_{vi}\, (v_i(t) - b) + \mathcal{F}_v(h,v_i(t)) +
  A_{vi}\;\xi_v(t) \nonumber \\
  \frac{\delta a_i(t)}{\delta t} &=& - \gamma_{ai} \, (a_i(t) - d) + \mathcal{F}_a(h,a_i(t)) + 
  A_{ai}\,\xi_a(t)
  \label{eq:va-basic}
\end{eqnarray}
The set of equations (\ref{eq:va-basic}) have three terms in the right-hand side: a exponential relaxation towards $[b, d]$  as a decay with parameters $\gamma_{vi}$ and $\gamma_{ai}$; a stochastic component  of amplitudes $A_{vi}$ and $ A_{ai}$  with random numbers $\xi_{v}(t)$, $\xi_{a}(t)$ drawn from a given distribution of white noise; and a component of external influences described by the functions $\mathcal{F}_v$, $\mathcal{F}_a$. 
These two functions formalize how the perception of online content in the field $h$ changes valence and arousal, where valence depends on the sign of $h$ and arousal depends on the absolute value of $h$. This way, the arousal dynamics respond to the overall presence of emotional content in the field, while valence depends on the polarity of the online interaction.
The functions $\mathcal{F}_v$, $\mathcal{F}_a$ are defined as:
\begin{eqnarray}
\mathcal{F}_v(h,v_i(t)) = h \times \left ( \sum_{k=0}^{3} b_k v_i(t)^k \right )= h * \left (b_0 + b_1 v_i(t) + b_2 v_i(t)^2  + b_3 v_i(t)^3 \right ) \nonumber \\
 \mathcal{F}_a(h,a_i(t)) =  |h| \times \left ( \sum_{k=0}^{3} d_k a_i(t)^k \right ) = |h| * \left( d_0 + d_1 a_i(t) + d_2 a_i(t)^2  + d_3 a_i(t)^3 \right ) 
  \label{eq:fva}
\end{eqnarray}
where $b_j$ and $d_j$ are parameters that define the changes in valence and arousal induced by different values of $h$. 
$\mathcal{F}_v$ depends on $h$, while $\mathcal{F}_a$ does on the absolute value $|h|$, modeling the excitability of arousal with emotional content independently of its polarity.
The dynamics of the field $h$ can be quantified through aggregates of the emotions in the messages of a discussion \cite{Garas2012,Garcia2011}, and are influenced by the emotional content of the expression of the agents, $s_i(t)$. 
This expression depends on the emotional state of the agent, following \emph{expression rules} and being triggered by states of high arousal. When produced, the expressed text carries a polarity that depends on the valence of the agent. We formalize this through the equation:
\begin{equation}
  \label{s-i}
  s_i(t) = f_s(v_i(t))\, \Theta[a_{i}(t) - \tau_i]
\end{equation}
where the expression of the agent has a value that is a function $f_s(v_i(t))$
of its valence. The expression $s_i(t)$ is activated  if the arousal reaches
a threshold value $\tau_i$, which is captured by the Heaviside step function
$\Theta[x]$.  The \emph{feedback of expression} produces an instant decrease in the arousal of the agent as an additional relaxation after writing a message.

In the following, we present an analysis of empirical data that explores emotion eigendynamics, perception dynamics, and rules and feedback of emotional expression through posts. 
Our results are based on a set of experiments that monitor the emotional states of their
participants in different ways.   
We present data from three experiments in which participants were exposed to online emotional content, including reading about emotional experiences and writing about highly emotional topics themselves. 
These experiments, explained more in detail in the Methods section, were designed to test the dynamics of online emotional interaction, and to provide data in the form of subjective assessment of emotions. 
We investigate emotion dynamics within our agent-based framework, estimating the values of the parameters that drive individual emotions.  
Due to the limited size of experimental data, we restrict our analysis to general emotion dynamics, ignoring individual differences. In the following, we drop the subindex $i$ of our dynamics, leaving the analysis of the differences emotion dynamics across individuals for future research.

\section{Methods}

\subsection{Experiment design}

\begin{itemize} 

\item \textbf{Study 1.}  
In the first experiment, 91 students from Jacobs University (54 female; mean age = 20.58; \textit{SD} = 2.34) read 20 threads  selected from various public online discussion forums such as BBC message boards for news and religion, as well as forums addressing more personal topics (e.g., lovingyou.com, femalesneakerfiend.com, loveshack.org) to cover a wide range of emotional topics ranging from contentious political discussions to questions about relationships, love, and heart-ache. All threads were drawn from a larger sample, and pre-categorized by 3 psychologists to contain text expressing emotions with certain polarity  (9 negative, 9 positive, 2 neutral). Stimulus valence (positive, neutral, negative) was varied as a within-subjects factor. The experiment was conducted online using Questback EFS survey (www.unipark.de), where the participants read the threads in randomized sequence, one post per page, at home on their own computers. Participants received a link to the experiment, were provided with a description of the study, and  generated an anonymous code to receive their compensation with course credit or 6 Euro. 
SI Figure 1 shows an example of a post as seen by the participants that was preselected as part of a negative thread taken from a real BBC forum discussion. 

After reading the posts of a thread, participants provided subjective reports of their emotions on three 7-point Likert scales to assess the subjective emotional response. With view towards the requirements of the repeated measurement situation (10 threads), we aimed to obtain sufficiently reliable measures while the relatively weak and fleeting emotional states that can be elicited by forum posts could still be expected to be present. For this purpose, e.g., even the two relatively short 10-item scales of the Positive and Negative Affect Schedule (PANAS, {Watson1988}) would have been too long and not sufficiently focused on the immediate emotional response to an individual emotional stimulus such as a forum thread. Instead, our measurement situation was much more comparable to experimental designs assessing emotional responses to emotional images, sounds, or words in the laboratory \cite{Lang1993,Larsen2003}. These types of designs typically measure valence, arousal, and sometimes dominance (power) via single-item graphical scales such as the Self-Assessment Manikin (SAM) \cite{Bradley1994}. Single-item Likert-type measures, if well phrased and designed, are not necessarily worse than multiple-item scales \cite{Gardner1998}. For example the SAM has been the basis for the validation of the extremely well cited International Affective Picture System (IAPS) \cite{Lang2008}. However, the graphical version of the SAM requires some additional instructions and explanations that are, ideally, delivered by an experimenter who can respond to questions. In addition, valence has recently been increasingly discussed as a potentially two-dimensional construct allowing co-activation of both positive and negative feeling states, such as in the case of bittersweet mixed emotions \cite{Larsen2011}. 

We assessed valence via two separate Likert scales, as well as arousal via a third scale comparable to the phrasing of the SAM. However, no systematic evidence for mixed-valence emotions was observed. Cronbach's $\alpha$=.88 across all positive and the inverse of all negative valence judgments suggested sufficient overlap between both items to justify integration of both scales into a single measure by subtracting the value of negative affect from the value of positive affect and mapping the result to the scale of  $[-1,1]$. The third Likert scale measured the degree of excitation experienced by the participant, which we likewise rescaled into our measurement of arousal in the scale $[-1,1]$. Finally, to obtain additional data beyond the strictly emotional response, participants answered a short set of 4 appraisal-related questions in response to each thread (perceived interest, relevance, wish to continue, wish to participate). We rescaled these answers to scales of $[0,1]$, measuring in particular the probability of participation.  The subjective assessments of emotions of this study are useful to study \emph{eigendynamics} and \emph{perception dynamics}, and the answers to the questionnaires provide data on how users decide when to create posts, following \emph{expression rules}.

\item \textbf{Study 2.} 
This experiment was an equivalent of study 1 in a controlled experimental setup. To further improve the validity of the scales in comparison to the online assessment, and to reduce unwanted anchoring and sequence effects that might occur when items are presented on-block in a numbered grid, each scale was presented individually on the screen with input provided via a 7-button response-pad without numbered keys (Cedrus RB-730).
The 7 threads used for this experiment are a subset of those for study 1 (3 positive, 3 negative, 1 neutral). Again, the emotional responses were assessed using 2 items for valence (1: "not at all positive" to "very positive"; 2: "not at all negative" to "very negative") and 1 item for arousal ("very calm" to "very excited"), however this time using the standardized response pad in a laboratory situation instead of the keyboard at home. Study 2 added data on 53 participants (21 female; mean age = 21.91; \textit{SD} = 3.74), which in combination with Study 1, resulted in two independent experiments which studied eigendynamics, perception dynamics, and expression rules. All participants were compensated with 6 Euro for this study. 

\item \textbf{Study 3.} 
In this laboratory experiment, 65 participants (30 female; mean age = 20.4; \textit{SD} = 1.9) were asked to write contributions to positive and negative emotional topics either in the form of replies or as initiators of new forum threads. Topics were presented and responses were collected using Medialab (Empirisoft). In a within-subjects design, all participants were instructed to respond to a selection of 5 short positive forum posts, a selection of 5 negative posts, and to initiate 1 positive topic as well as 1 negative topic. This procedure resulted in the production of 4 forum posts per participant. When asked to respond to a forum post, the ability to choose from among a selection of pretested topics aimed to increase personal relevance of the chosen topic's content, and thus the emotional significance, while making it easy for participants to respond. When asked to initiate a new topic with a first post of their own, participants received direct on-screen instructions to write about a positive vs. negative topic that they liked vs. disliked, felt good vs. bad about, and that they would be willing to share with others. Thus, while participants knew that their input to both types of forum discussions would not be posted online, they were asked to "imagine that the text you write will be posted to an online forum, a newsgroup, or an open chat". We expected that some participants might find it easier to be asked to respond to a topic of their choice while others might have a greater preference for a more open discussion topic that they could define on their own, and this design aimed to facilitate the generation of overall sufficiently long responses from all subjects. Nevertheless, participants were free to write as much or as little as they liked.
The participants provided subjective assessments of their emotions before and after writing the posts, which we will use to study the \emph{feedback of expression} into emotions.

\end{itemize}

Conforming with standard ethical guidelines, participation in all three of the above studies was entirely voluntary, and participants were informed that they could quit participating in the study at any point in time without any negative consequences. Participants provided informed consent at the beginning of a study, and all participants were fully debriefed at the end of each of the  studies in this research. Furthermore, to avoid any negative impact on participant's emotional well-being during or after taking part in the study, the study design provided a balancing positive emotional stimulus for each negative stimulus that was presented, thus preventing any buildup of negative emotional states. In the laboratory studies, participants were furthermore given the opportunity to ask questions to the experimenter who ensured that there were no potentially lingering negative effects of the negative forum threads at the end of the study.

\subsection{Regression models of perception dynamics}

Using an event timescale, we aggregate the data in a set of changes per
time unit $\Delta v(t)/\Delta t$ and $\Delta a(t)/\Delta t$, or valence and arousal "velocity". 
We then analyze their dependence on the instant value before reading the thread $v(t)$, $a(t)$, and the emotional charge of the thread ($h$). 
Instead of fitting the solution of the equations of emotion dynamics of equation \ref{eq:va-basic} to the data, we used a regression model to estimate the changes in the variables as a combination of a linear and  a stochastic component.  
This has the advantage  that we do not need an explicit solution of equations \ref{eq:va-basic}, allowing us to test nonlinear relationships like the ones shown in equation \ref{eq:fva}.

Our experimental design defines three cases of $h$, corresponding to the fields generated by  positive ($h=+1$), negative ($h=-1$) and neutral ($h=0$) threads. 
We reformulate the continuous dynamics of equations \ref{eq:va-basic} and \ref{eq:fva} as discrete changes due to a combination of a relaxation force and an stimulus-dependent influence for valence:
\begin{equation}
\frac{\Delta v(t)}{\Delta t} = - \gamma_v \left ( v(t) - b \right) +  h * \left (b_0 + b_1 v(t) + b_2 v(t)^2  + b_3 v(t)^3 \right ) + A_v \epsilon
\label{eq:hVest}
\end{equation}
and the dynamics of arousal as dependent on the absolute value of $h$:
\begin{equation}
\frac{\Delta a(t)}{\Delta t} = - \gamma_a \left ( a(t) - d \right) +  |h| * \left (d_0 + d_1 a(t) + d_2 a(t)^2  + d_3 a(t)^3 \right ) + A_a \epsilon
\label{eq:hAest}
\end{equation}

First, to test the relevance of each term of the equations and avoid
overfitting,  we search for the best parameter subset through a maximum
likelihood procedure \cite{Venables2002} that minimizes  the  Akaike
Information Criterion \cite{Akaike1981}. Second, we fit the most informative
parameter subset of equations \ref{eq:hVest} and \ref{eq:hAest} through the
non-linear least squares method \cite{Bates1988} (more details in SI). Third, we compute an empirical estimator $\widehat p$ for each relevant parameter, finding the
posterior distribution of $\widehat p$ in a Bayesian model with normally
distributed priors \cite{Gelman2008} and $10.000$ simulated samples.

\subsection{Sentiment analysis}

To test how emotions are encoded in text through function $f_s$, we apply two
sentiment analysis tools to  the 182 posts produced in Study 3. First, we
apply SentiStrength \cite{Thelwall2010} a state of the art sentiment tool that
quantifies positive and negative content independently. SentiStrength is among
the best performing sentiment analysis tools in benchmark tests of social
media posts and comments \cite{Goncalves2013}, and has been use to study blog
posts \cite{Chmiel2011}, chatroom messages \cite{Garas2012}, YouTube comments
\cite{Garcia2012} and microblog posts \cite{Thelwall2011}. SentiStrength
provides two scales of positive, from 1 to 5, and negative sentiment, from -1
to -5. We classify a post as positive if its positive score is 3 or higher,
and negative if its negative score is -3 or lower, following the methods of
previous research \cite{Thelwall2010,Thelwall2011,Garcia2012}.

The second tool we apply is QDAP (Quantitative Discourse Analysis Package)
\cite{Rinker2013}, using the Opinion mining lexicon \cite{Hu2004}. QDAP
matches words against the lexicon and estimates a range of polarity between -1
and 1 for each sentence in the post.  We classify a post as negative if the
minimum value of polarity among the sentences in the post is below -0.25, and
positive if the maximum value among sentences is above 0.25. Note that, in
line with SentiStrength, this method can detect simultaneous positive and
negative content in a post.

\section{Results}

\subsection{Eigendynamics and perception dynamics}
\label{sec:perception}

In experiments 1 and  2, the order of the threads read by the participants was randomly determined, keeping the post ordering inside the thread but randomizing when each thread is presented.  
Participants were asked to provide their emotional reports between threads in order to obtain as accurate measures as possible without interfering with the task \cite{Robinson2002}.
This sequence of tasks allows us to reconstruct the changes induced in valence and arousal after reading threads with different emotional content, in order to know how online discussions expressed in the stimuli influence the emotional state of a user. 
These influences compose the \emph{perception dynamics} of online emotional interaction, and coexist with an \emph{eigendynamics} that are independent of the perceived content.

We formulate regression models (explained in the Methods section) to test the dynamics expressed in equations \ref{eq:va-basic} and \ref{eq:fva}. Using a maximum likelihood criterion, we find the most informative parameter subsets that explain the empirical changes in emotions, and compute the posterior distribution of each parameter in a Bayesian model with normally distributed priors \cite{Gelman2008}.

\subsubsection{Valence dynamics}

\begin{figure}[!ht]
\begin{center}
\includegraphics[width=0.9\textwidth]{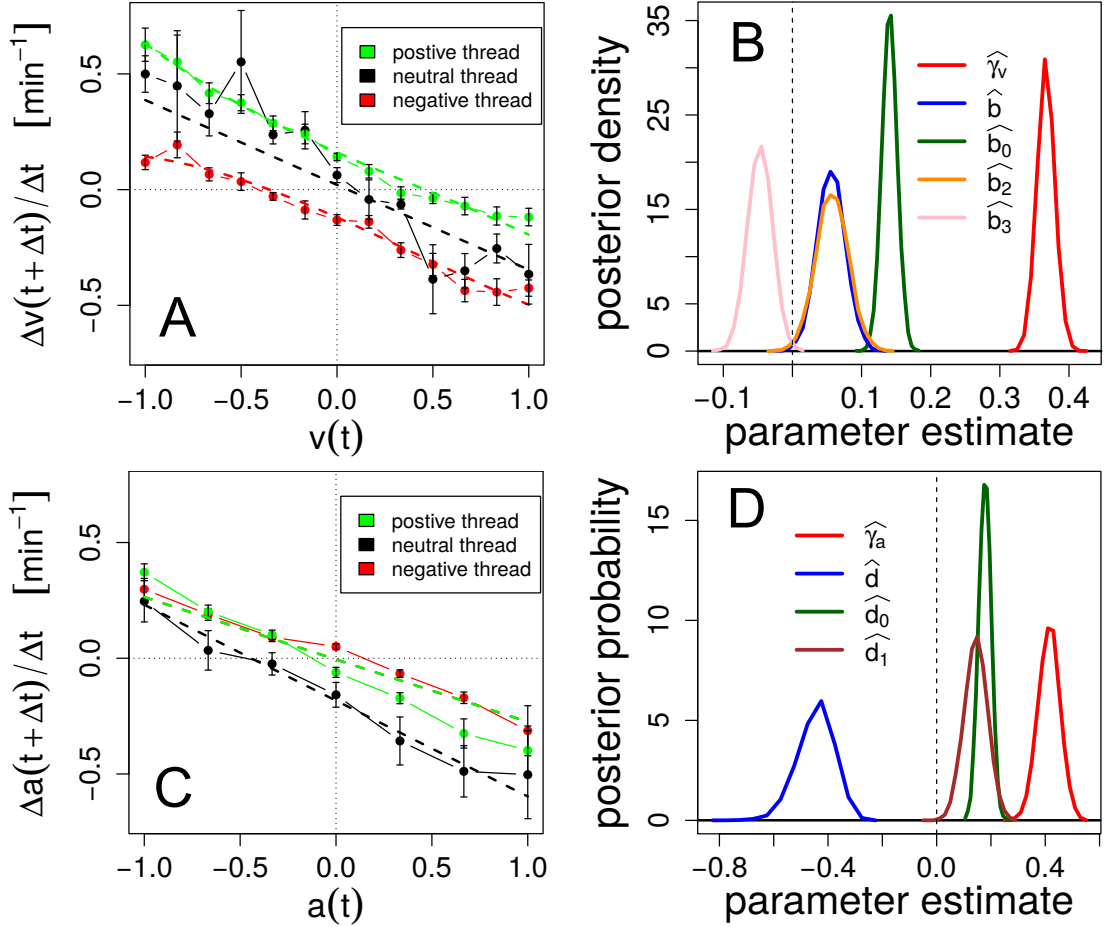}
\end{center}
\caption{ Mean change in valence (A) and arousal (C) per time unit while reading threads versus the previous reported value. 
Error bars show standard error, dashed lines show the changes predicted by the model.
B,D) Posterior density function of relevant parameters in valence (B) and arousal (D) dynamics over 10.000 simulations, binned with Sturges' formula \cite{Sturges1926}. \label{fig:VReadingChanges} } \end{figure}

For the case of $v(t)$ the maximum likelihood procedure detects that the
equation of valence has significant terms up to the third order, with the
exception of $b_1$, which can be considered as $0$ (details in SI).
The results of the nonlinear regression are shown in Table \ref{tab:VregAll}. The
eigendynamics captured by $\gamma_v$ and $b$ describe a
fast relaxation process towards a small valence baseline ($0.056$), in which
valence decreases by more than 30\% per minute. This can be observed in the
valence change function for neutral threads in Figure
\ref{fig:VReadingChanges} A, which takes the form of a negative slope that
crosses the horizontal axis close to $v=0$.

The strongest effect of $h$ in the valence
is through $b_0$, which shifts the valence by a constant factor of
$0.14$ on the direction of the emotional charge of the thread. The nonlinear
terms $b_2$ and $b_3$ are significant but small in
magnitude, showing  corrections only close to the limits of strong positive and
negative valence.

\begin{table}[h]
\centering
\begin{tabular}{ c | c  c  c  c  c| c  c c }
parameter 	& $\gamma_v$		& $b$   			& $b_0$  			& $b_2$   		& $b_3$  		  
& $R^2$  & $N$  & $R^2(\xi_v)$  \\ \hline 
estimate    & $0.367^{***}$	& $0.056^{**}$	& $0.14^{***}$ 	& $0.057^{*}$ & $-0.047^{**}$ & $0.52$ & $1271$ & $0.85$
\end{tabular}
\caption{Parameter estimations of equation \ref{eq:hVest}, $R^2$ value of
nonlinear least squares, and  $R^2(\xi_v)$ of a normal fit to the residuals of
the regression.    $^{*} p<0.01$, $^{**} p<0.001$,$^{***} p<10^{-10}$
\label{tab:VregAll}}
\end{table}

The posterior distribution of each parameter estimate is shown Figure  \ref{fig:VReadingChanges} B, showing the location and
uncertainty of each parameter estimate given the empirical data.   These
results lend strong evidence for our hypotheses drawn from the Cyberemotions
modeling framework. All parameters can be considered different from $0$, and
their effect sizes reveal the existence of  relaxation eigendynamics and a
linear  influence of $h$ with nonlinear corrections. To further test the
robustness of this result, we evaluated  the interaction of the experimental
setup in our statistical analysis (see SI). We find no significant effect in
any estimator with the exception of $\gamma_v$, which differed on a
total of $0.06$ across experiments, less than $20\%$ of its point estimate. In
addition, the regression results suggest an important level of predictability
of the valence, with an $R^2$ value above $0.5$. The assumption that error terms
are normally distributed is not far from reality, as a normal fit to the
residuals through the method of moments gives $R^2(\xi_v)=0.85$ and a Shapiro-Wilk statistic of $0.8762$.

\subsubsection{Arousal dynamics}

For the case of arousal, the maximum likelihood method reveals that the
nonlinear terms $d_2$ and $d_3$ are not significant. The
resulting model is summarized in Table \ref{tab:AregAll}, where the arousal
eigendynamics show a strong relaxation toward a negative arousal baseline of
$d=-0.442$ at a rate of decrease above 40\% per minute. This can be
observed on Figure \ref{fig:VReadingChanges} C, in which the
function of arousal changes for neutral threads crosses the horizontal axis at
a negative arousal value. The other two parameters of arousal dynamics,
$d_0$ and $d_1$, show that online content has a positive
effect on arousal regardless of the polarity of its emotional charge, and that
the attraction speeds up with a factor close to $0.15$. 

\begin{table}[!ht]
\centering
\begin{tabular}{ c | c  c  c  c | c  c c }
parameter 	& $\gamma_a$		& $d$   			& $d_0$  			& $d_1$   	 & $R^2$  & $N$  & $R^2(\xi_a)$  \\ \hline 
estimate    & $0.414^{***}$	& $-0.442^{***}$	& $0.178^{***}$ 	& $0.14469^{**}$ & $0.28$ & $1271$ & $0.78$
\end{tabular}
\caption{Parameter estimations of equation \ref{eq:hAest}, $R^2$ value of
nonlinear least squares, and  $R^2(\xi_v)$ of a normal fit to the residuals of
the regression.    $^{*} p<0.01$, $^{**} p<0.001$,$^{***} p<10^{-10}$
\label{tab:AregAll}}
\end{table}

The posterior distribution of each arousal parameter is shown on Figure  \ref{fig:VReadingChanges} D.  As hypothesized in the Cyberemotions framework,  a relaxation component toward negative arousal coexists with a force that increases arousal regardless of the sign of $h$, as captured by  $d_0$ and
$d_1$.  
The $R^2$ value of the arousal model is lower than for the case of valence, and the quality of the assumption of normal error is also lower.
This suggests the existence of additional terms  beyond those included in equation \ref{eq:fva}.
We tested if we should include a linear dependence on $h$  in addition to its absolute value on an extended model.
We found negligible linear effects of $h$ (more details in SI), supporting the assumption that arousal depends on $h$ but not on its sign.
Furthermore, we tested the effect of the experimental setup as
we did with the valence dynamics, finding that all parameters  remain
significant when controlling for the experimental conditions.  The control
parameters for the experimental setup were not significant for any parameter
but $d_0$, which showed an attenuated effect in Study 2, yet remained
significant and sizable (more details in SI).

\subsubsection{Simulating perception and eigendynamics}

One of the advantages of an agent-based computational approach is that allows us to
implement computational equivalents to compare empirical and simulation
results. Not only can we provide a quantitative explanation for the observed
data, but we can reproduce its behavior in simulations, or even apply it in
the field of affective computing \cite{Rank2013}.  We simulated emotional
agents using the parameter estimations of Table \ref{tab:VregAll} and Table
\ref{tab:AregAll}, computing their changes in valence and arousal when exposed
to values of $h$ corresponding to the emotional charge of threads. 
\begin{figure}[ht]
\centering
\includegraphics[width=0.95\textwidth]{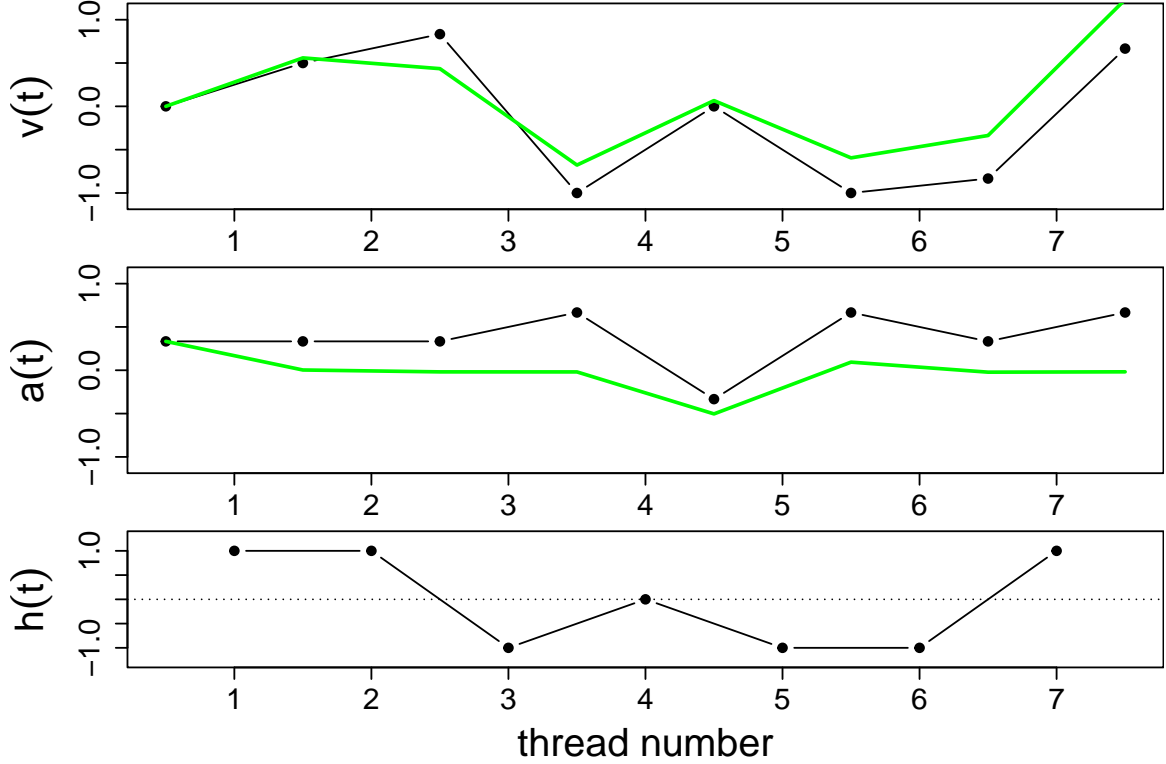}
\caption{Data-driven simulation of perception dynamics in an experiment. 
The top panel shows valence, the middle arousal, and the bottom shows the emotional charge of the thread. 
Black lines show the empirical data of a participant of Study 2, green lines show the results of a simulated agent starting from the same emotional state. \label{fig:Simulation} } \end{figure}
Figure \ref{fig:Simulation} shows an example of a simulation of the model
versus the sequence of responses and emotional charges of threads for a
participant of Study 2. 
This illustrates the dynamics of an individual, rather than analyzing the average
response as shown in the regression results of Table \ref{tab:AregAll}.
The simulated and empirical data are similar, in
particular with respect to the sign of movements of valence and arousal after
each thread type. This kind of computational model serves as a building block
to estimate, predict, and reproduce emotional dynamics under online
interaction.

\subsection{Expression dynamics}

\subsubsection{Production rules}

\begin{figure}[h]
\centering
\includegraphics[width=0.98\textwidth]{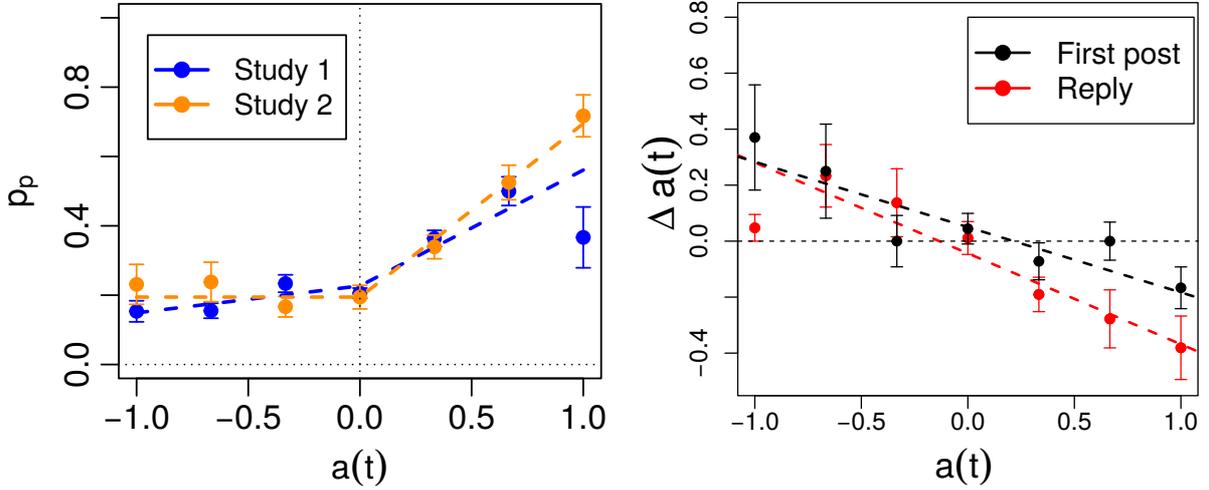}
\caption{Left: Mean reported
  participation intention given experience arousal in Studies 1 and 2. Right:
  Change in arousal when producing first posts and comments in Study 3. Error bars show standard
  error. Dashed lines show MARS fits on the left, and linear regression results on the right.
  \label{fig:Participate} }
\end{figure}

In Studies 1 and 2, participants were asked about their intention to
participate in a given discussion. One of the assumptions of the Cyberemotions
framework is that arousal is the driving force behind user participation in
online discussions.  This assumption is formulated in equation \ref{s-i}, as a
threshold function depending on arousal. We test this assumption through a
hypothesis of a  linear relation between participation tendency and
experienced arousal, which starts from a given point of arousal:
\begin{equation}
\label{eq:mars}
p(t) = p_0 + \alpha * a(t) * \Theta[a(t) - \tau]
\end{equation}
We fit the above equation through  the method of Multivariate Adaptive
Regression Splines (MARS) \cite{Friedman1991}, finding  the best fitting
values for  $p_0$, $\alpha$, and $\tau$. Table \ref{tab:Amars} shows the
result of the fit, and the left panel of  Figure \ref{fig:Participate} shows
the relation for both studies.  The intention to participate is heavily
influenced by arousal when it is above $0$. Below that level, participants had
a ground tendency  close to $0.2$, but for positive arousal the participation
tendency grew with a significant trend above $0.4$ (see SI for more details).
This pattern is similar in both studies, with growing participation for
arousal above $0$. The differences between experiments is not relevant in the
parameter estimates as  formulated in equation \ref{eq:mars}, and only some
deviation could be found for very high arousals when introducing an additional
breakpoint (see SI). We also tested a possible relation of participation
tendency with valence, finding a weaker relation with a small increase for
very positive valence, as explained in the SI.

\begin{table}[h]
\centering
\begin{tabular}{ c | c  c  c  | c }
parameter & $p_0$   & $\alpha$  & $\tau$ & $R^2$ \\ \hline
estimate  & $0.199$ & $0.438$   & $0$    & $0.14$
\end{tabular}
\caption[Participation MARS results]{Parameter estimations of equation \ref{eq:mars} and $R^2$
\label{tab:Amars}}
\end{table}

\subsubsection{Expression function} 

The interactive setup of Study 3 allows us to test the dependence between
online expression through text and subjective emotions, formalized in equation
\ref{s-i}. Our hypothesis is that the emotional content measured through
text, $s_i$, is a function $f_s(v_i(t))$ increasing with valence but not with
arousal. We test this hypothesis by applying two state of the art sentiment
analysis techniques to the text produced by participants in Study 3,
quantifying whether a text contains positive and/or negative content
simultaneously (more details in Methods). This way, for each post $p$ we count
with two variables $pos_p$ and $neg_p$ that take the value $1$ if the text
contains positive and negative content respectively, and $0$ otherwise. On the
side of subjective reports of emotions, we count with the subjective valence
and arousal reported after writing the post, $v_p$ and $a_p$.

\begin{figure}[ht]
\centering
\includegraphics[width=0.95\textwidth]{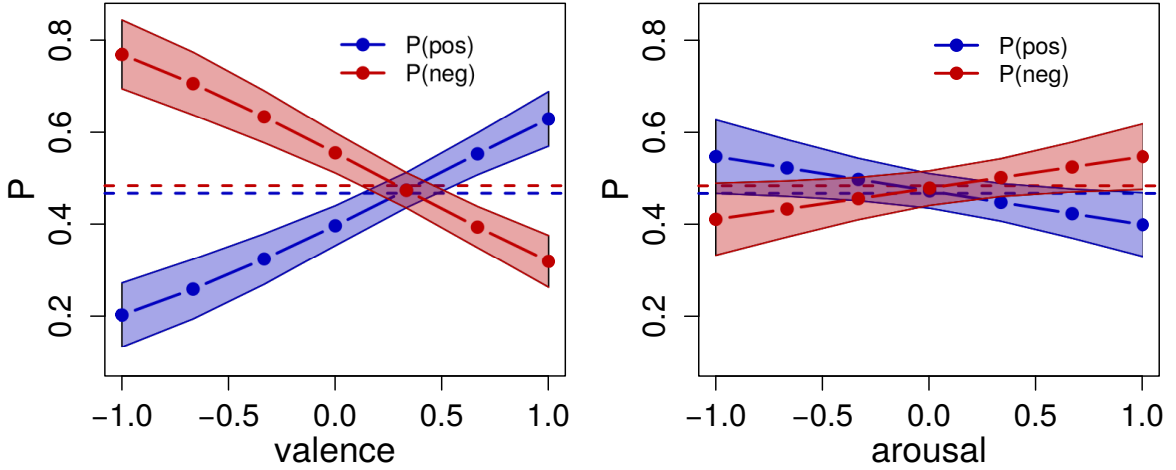}
\caption{Results of logistic regression of post positive and negative content measured with
SentiStrength as a function of valence and arousal. Error bars show standard errors of the estimate of
the probability of being positive or negative.\label{fig:Expression} } \end{figure}

We test the relation between the content of posts and the subjective
experience of participants through logistic regression models
\cite{Crawley2014}, formulated as $logit(P(pos))=p_0+\alpha_v*v$ and
$logit(P(pos))=p_0+\alpha_a*a$ for positive content and similarly for negative
content. Figure \ref{fig:Expression} shows the result of the models for
positive and negative content in the post as estimated with SentiStrength. We
find a significant influence of valence on the probability that a post is
positive and negative, as reported in detail in Table
\ref{tab:Expressionregs}. In addition, we do not find statistical evidence of
the role of arousal, in line with our assumption of $f_s$ as a function of the
valence only. These results are robust when using an alternative sentiment
analysis technique (see SI).

\begin{table}[!ht]
\centering

\begin{tabular}{ c | c  c  c  | c c c }
model & $p_0$ 					& $\alpha_v$ & $\alpha_a$ & Log Likelihood & Deviance  & $N$   \\ \hline 
pos   & $\mathbf{-0.4203}^{*}$ 	& $\mathbf{0.9462}^{**}$  & &  -120.3680 &  240.7359 & 182 \\
neg   & $0.2194$                & $\mathbf{-0.9777}^{**}$ & &  -120.2017 & 240.4034 & 182 \\
pos   & $-0.1110$  &  & $-0.0865$ & -125.1264 & 250.2527 & 182 \\
neg   & $-0.2990$  &  & $0.2743$ & -125.5221 & 251.0442 & 182 \\

\end{tabular}
\caption{Logistic regression results for positive and negative post content as a function of reported valence and arousal   $^{*} p<0.01$, $^{**} p<0.001$,$^{***} p<10^{-3}$
\label{tab:Expressionregs}}
\end{table}

\subsubsection{Feedback of expression}

The experimental setup of Study 3 focused on the production of posts, with
subjective assessment of emotions before and after writing the post. In a
similar way as we did for the changes after reading a thread, we calculated
the feedback of expression in the emotional state of an individual
after writing a post. In the Cyberemotions framework,
we hypothesized a reset to $a=0$ after a user posted a message, as an implicit
regulation mechanism after emotional expression ("auto regulation" \cite{Kappas2011,Kappas2013b}). The change in arousal
depending on previous arousal for first posts and replies is shown on the
right panel of  Figure \ref{fig:Participate}. For both types of production,
negative arousals tend to increase and positive arousals to decrease toward a
neutral value. We validated this observation through a linear regression, and
found no significant difference between writing a first post or a reply (see
SI). In addition, a similar effect was present in valence, but only for the
case of replies. Negative valence experienced an increase after replying in a conversation, as explained more in detail in the SI.

The changes in arousal after participation do not show a reset to a value of zero
after writing a post, which was the initial assumption in the Cyberemotions
framework. On the contrary, humans seem to experience a decrease on their
arousal that does not necessarily reset it to the neutral value, which still
leaves room for further activity even in the absence of interaction. This
difference in posting dynamics should be taken into account in future models
within our framework, but the causality between participation and arousal is
supported by these results. When an Internet user perceives emotional content, her arousal increases, leading to higher chances of participating
in the discussion. This participation induces an instant decrease in
arousal, which, in combination with the internal relaxation of the
arousal, would decrease the probability of further participation.
If other users create more emotional content (of any valence sign), the
arousal of the agent would increase again, leading to the coupling of user
behavior that explains collective behavior in online communities.

\section{Discussion}

The work presented here shows how online communication influences the emotions
of an Internet user, and how these emotions change over time. Our work aimed
at testing the assumptions of the Cyberemotions agent-based modeling
framework. Analyzing subjective assessments of emotions, we found strong
support for the presence of an exponential relaxation towards a ground state.
In terms of emotional content in online discussions, we found that valence
changes according to the polarity of threads, and that threads with emotional
content lead to higher arousal than threads with neutral content. The agents
in our framework express their emotions when their arousal reaches a
particular threshold. We verify this by inspecting the dependence between
reported arousal and the intention to participate in a conversation, which
reveals that such participation tendency increases linearly when the arousal
is beyond a threshold value. Another assumption of our modeling framework was
the effect of writing a post on arousal, which we hypothesized as an instant
decrease of arousal. Our empirical results indicate the existence of a
decrease after replying to posts in an online discussion, when participants
were in high arousal states before interaction.

In general, our results are consistent across experimental conditions in
studies 1 and 2, but some discrepancies are present for arousal. For example,
it is likely that the unfamiliar laboratory situation in study 2 as such may
have elevated the average arousal level of participants. This would be
consistent with notions of prior psychological research on the possibility of
misattribution and transfer of arousal from another source of activation which
may occur under certain conditions \cite{Zillmann1974,Cummins2012}.   In study
2, participants were more likely to feel more observed as well as more engaged
in the task than in study 1, which they completed at home. Intentional
regulation attempts may have been facilitated by the more public social
context of study 2, as well as high levels of excitation in response to some
of the threads \cite{Kappas2011}. The present research required very
brief measures to avoid subject fatigue and dissipation of the
emotional impact of having read the short threads, which limited the number of
items that could be posed repeatedly. Compared to single-item Likert-scales, a
two-dimensional single-item measure  such as the Affect Grid
\cite{Russell1989} could have been even briefer, once sufficiently explained.
However, this type of measure still faces conceptual issues \cite{Gray2007},
and specifically the good validity of the Affect Grid reported by the original
authors \cite{Russell1989} has been found to be only moderate by subsequent
research \cite{Killgore1998}.

In psychology, there is increasing awareness of the social nature of
emotional-contagion processes \cite{Kleef2009}, as well as a clear interest in
the psychological consequences of the social sharing of emotions
\cite{Curci2012,Nils2012}. However, large-scale research on emotions in
cyberspace is still a very recent development \cite{Kappas2013},  and our work
is the first one to show an integration of modeling with  research in the laboratory. At
the same time, there is an increasing sensitivity within the psychology about the use of data from large social networks such as Facebook,
concerning established principles of informed consent and the opportunity to
opt out of participation in any kind of experiment. For example, a recent
article \cite{Kramer2014} has led to a statement of editorial concern
regarding these aspects of personal privacy, and the young field of large-scale analysis  online emotions  is likely to be faced with continued
questions in this regard. In contrast, in the present paper we have gone a
different route to test assumptions of large scale modeling of collective
emotions by means of taking some of the central processes into the laboratory.
As a result we have revealed much stronger and salient effects than the
previously reported linguistic cues \cite{Kramer2014},  showing that emotional interaction online goes well beyond subtle contagion processes.

From a psychological perspective, our findings are generally consistent with
the assumptions of the Cyberemotions framework.  The relation between
participation and arousal above a threshold is of particular importance. This
links our present results to some classic findings in experimental psychology,
in particular with respect to the role of excitation on subsequent emotional
behavior \cite{Cantor1974,Cummins2012,Turnbull2002,Zillmann1974}. Thus, while
the concept of a carry over of arousal is not new to psychology, computational
modeling of these types of behaviors promise a more direct means of testing
the precise dynamics involved in these processes.  In addition, a
computational model supports the transfer and application of our
findings to large scale phenomena on the Internet involving collective
emotions. For example, at present, our data suggest that interventions aimed
at a change of arousal might a be promising approach to the calming the nerves
in a heated online discussion, in comparison to  valence-based interventions
that primarily aim at the polarity of an ongoing discussion. Enhancing
user interaction to create collective emotions is also relevant for the design
of automatic dialog systems \cite{Rank2013} and virtual human platforms
\cite{Ahn2012}.

The dynamics of emotional states during online interaction show that arousal
is driven towards negative values for non-emotional threads, which we did not
take into account in the initial models of our framework. In addition, we
found that replying to a post creates a valence increase for users that have a
negative valence before interaction, as a beneficial result of the
participation in an online discussion. Furthermore, our assumption of the
decrease in arousal after expression is hereby extended, as we find that such
a decrease does not reset the arousal to 0, but lowers it by some proportional
amount. Further experiments shall focus on how simultaneous positive and
negative emotional content influences the emotions of the readers of a thread.
Further work should study how subjective emotions are encoded in expressions,
which is included in our model through the function 
$f_s(v_i)$. 
Testing the way emotions are
encoded in a text heavily depends on the sentiment analysis tool used to
process the posts, for which particular experimental designs are necessary.
While still some of these hypotheses remain untested, the results reported here allow the design of better agent-based models of emotions and contribute to a better understanding of the emergence of collective emotional states on the Internet.

\bibliographystyle{plain} \bibliography{EmotionExperiments} 

\section*{Ethics statement}
Procedures used involved standard paradigms and were consistent with the rules of the American Psychological Association and the Declaration of Helsinki. All laboratory studies involved fully informed consent. Approval regarding ethical conduct was obtained from the external advisory board of the CYBEREMOTIONS project for all experiments.

\section*{Data accessibility}
The data and codes used to produce the results shown here can be access at\\
\texttt{http://github.com/dgarcia-eu/EmotionDynamics\_RSOS}

\section*{Competing interests}
The authors declare no competing interests.

\section*{Authors' contributions}
AK and DK designed and performed experiments; DG and FS designed agent-based model; DG 
carried out the statistical analyses; all authors participated on writing the article and gave final approval for publication

\section*{Acknowledgments}
The authors thank Simon Schweighofer for the useful feedback and discussions, and Elena Tsankova and Mathias Theunis for their help with collecting the empirical data for these experiments.

\section*{Funding statement}
This research has received funding from the European Community's Seventh Framework Programme FP7-ICT-2008-3 under grant agreement no 231323 (CYBEREMOTIONS). DG and FS received funding from the Swiss National Science Foundation (Grant CR21I1\_146499).

\end{document}